# Dielectric microparticles for enhanced optical imaging: a FDTD analysis of contrast and resolution

ARLEN R. BEKIROV[1,*], ZENGBO WANG[2], BORIS S. LUK'YANCHUK[1], AND ANDREY A. FEDYANIN[1]

[1] Faculty of Physics, Lomonosov Moscow State University, Moscow 119991, Russia
[2] School of Computer Science and Engineering, Bangor University, Bangor, Gwynedd, LL57 1UT, UK
*bekirovar@my.msu.ru

**Abstract:** This paper presents a comprehensive numerical analysis of super-resolution imaging using dielectric microparticles, employing the Finite-Difference Time-Domain (FDTD) method to elucidate the mechanisms that enable resolution enhancements beyond the diffraction limit. Our study demonstrates that dielectric microparticles can achieve a resolution on the order of 50 nm in the visible spectrum, surpassing traditional optical microscopy limits. By simulating the propagation of radiation through a microparticle-object system and generating optical images via a backward propagation technique, we reveal critical insights into how microparticles enhance image contrast and resolution. The study also explores the influence of various parameters, such as source coherence and particle-substrate interactions, on the image formation process. Our results not only validate the super-resolution capability of microparticle-assisted imaging but also provide a robust framework for further advancements in optical imaging technologies, with potential applications in fields requiring ultra-high-resolution visualization.

## 1. Introduction

The wave nature of light limits the size of objects that can be resolved using an optical microscope. According to the Abbe criterion, two point incoherent sources are indistinguishable if the distance between them is less than half of the wavelength $\lambda/2$ of the emitted radiation. It was shown [1] that by examining objects with the help of dielectric microparticles, one can resolve structures beyond the diffraction limit. Various theoretical approaches have been proposed to explain this phenomenon [2-13]. A detailed review of both experimental and theoretical results is presented in Ref. [14]. Despite these efforts, the underlying mechanisms that drive this super-resolution imaging remain incompletely understood, necessitating a deeper exploration. In order to explain the theoretical problems, it is necessary to consider in more detail the process of image formation by a microparticle.

This study begins by examining this process using geometric optics. While observing an object, for example, a point source, through a microsphere, the ray paths are refracted at the microparticle boundary, see Fig. 1. If the refractive index of the sphere is about 1.5, a virtual magnified image is formed. The distance between the two sources increases; however, it would be a mistake to say that such consideration explains overcoming the diffraction limit. The rays forming the virtual image are limited by a cone with an opening angle α, see Fig. 1, whose boundaries touch the surface of the sphere. As a result, the numerical aperture that forms the image is limited. The spot size d for a virtual image cannot be less than $\lambda/(2\sin\alpha)$. Thus, the apparent increase in image size due to magnification does not fully account for the observed super-resolution effects.

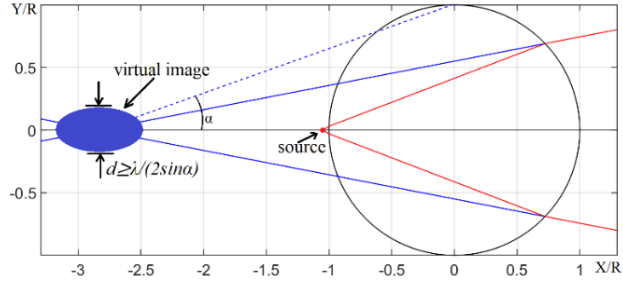

**Fig. 1**. Geometric path of rays from a point source. Red rays are the actual path of the rays, blue ones form a virtual image and are a continuation of the red ones during reverse propagation. The blue oval is the size of virtual images in the present of the microparticle. The image is significantly blurred due to the limited aperture. R is the radius of the particle.

This consideration was carried out from the standpoint of geometric optics, which has a limited scope of application. The condition for applicability is that the light wavelength is much smaller compared to the particle size. In experiments, the particle size ranges from several to tens of wavelengths [1, 11]. Given these limitations, it is advisable to also consider this problem from the standpoint of wave optics. The wave theory generally agrees with geometric considerations [5], except in cases of resonances. It was firstly demonstrated in Ref. [4] that exciting the whispering gallery modes with a dipole increases the resolution to $\lambda/4$, but requires a very special choice of the wavelength or the radius of the sphere. Typically, experiments show super-resolution using a broadband light source with microparticles of various sizes. Therefore, a valid theory should be independent on such resonance phenomena. Additionally, the image field in Ref. [4] exhibits significant side lobes around the focusing point, which were not observed experimentally.

The situation changes when the substrate is taken into account, as it introduces additional rays due to re-reflection of the "sphere-substrate" type. We have recently demonstrated [12] that a particle-on-substrate system allows magnification without blurring with a resolution of $\lambda/8$. However, modifications to the solution are necessary, such as changes in the source field or in the scattering matrix of the sphere. These modifications significantly impact the original solution. It is important to note that these are the only possible amendments to achieve such enhancement. Thus, despite the purely theoretical nature of the work, it revealed that the system's required properties to overcome the diffraction limit differ significantly from those previously considered. This insight suggests the need to explore more complex interactions, such as those between the microparticle and the object through light scattering. To address this, we propose a complete simulation of radiation propagation from an object through a microparticle and subsequent image formation. Due to the high computational demands of such simulations, we limit our study to the two-dimensional case.

## 2. Results

### 2.1 Halogen lamp source

We begin by simulating the results from Ref. [1]. The experiment employed an incoherent source, specifically, a halogen lamp. Importantly, the subsequent results remain unchanged even if a uniform spectrum from 425 nm to 675 nm is considered. To account for the spatial coherence of the field, we specified a source with a finite spatial size, which determines the coherence radius.

We employed the 2D FDTD method in TE geometry ($E_z=0$), as implemented in the Lumerical MODE package. A custom time signal with a specific spectrum was used to construct the source, allowing for the precise control of the simulation parameters. To generate the image, we applied the concept of radiation backpropagation. For this purpose, we utilized a 2D FDTD method implemented in custom Matlab code. To generate the image field, the radiation sources are time-reversed real fields, which then propagate in free space. The source can be specified based on a single component $H_z$:

$$H_z^{image\ source}(x, y = y_0, t) = H_z^{real\ field}(x, y = y_0, -t) \tag{1}$$

To record the real field, we used time monitors positioned above the microparticle. Both reflecting and transmitting imaging modes were investigated, with the calculation scheme presented in Fig. 2. In reflection mode, sources were placed above the microparticle. To exclude the field from the sources when the field is reversed in time, we used directed sources, as implemented in Lumerical. In this configuration, the source field extends in only one direction indicated by the black arrows in Fig. 2. For transmission mode, the source was placed below the sample. It is important to note that the resulting image field does not depend on the location of the reversal plane $y_0$, provided that the distance to this plane is much larger than $\lambda$. In our calculations, this distance was greater than 5 μm. To better match the experimental data, we used a spatial coherence radius of 600 nm for the source. Further details on the image field generation algorithm are provided in the Materials and Methods section.

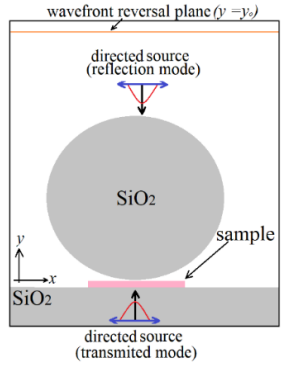

**Fig. 2.** The calculation scheme.

Firstly, we simulated super-resolution for slits in a metallic screen in transmitted light mode. The sample consists of a 30 nm thick perfect electric conductor (PEC) screen coated on a glass substrate, with four 360 nm wide slits spaced 130 nm apart. A glass microparticle with a radius of 2.37 μm is positioned above the slits, as shown in Fig. 3(c). We simulate four slits offset from the center of the sphere to ensure that the maxima in the image field accurately identify the location and the number of slits. The source is located inside the glass substrate, 1.25 μm below the slits.

The simulation results are presented in Fig. 3. The microparticle enables the distinction of slits, which are indistinguishable in free space. We plotted the square of the average intensity, $\left(I^{image}\right)^2$, which does not affect the resolution but provides better contrast. Some maxima in Fig. 3(b) arise due to partially coherent effects.

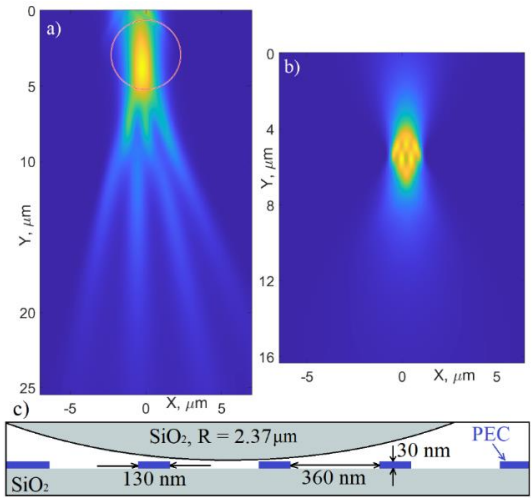

**Fig. 3.** Comparison of the image field in transmission mode for the slits in a metallic screen for free space and a microsphere. (a, b) Time-averaged image fields with and without a microparticle, respectively; the red circle denotes the boundary of the particle. Comparing distributions (a) and (b), we can conclude that the slits are indistinguishable without the presence of the microparticle. For better contrast, we used $(I^{image})^2$. The maxima in (b) are associated partial coherence and disappear at a smaller coherence radius, although the slits become distinguishable. (c) General calculation scheme: four 360 nm slits with 130 nm spacing in a 30 nm thick perfect conductive screen placed on a glass substrate. The source is located inside the substrate, 1.25 μm below the slits, and a glass sphere with a radius of 2.37 μm is positioned above the slits.

Next, we consider super-resolution for a Blu-ray DVD disk in reflection mode. The DVD disk is represented as perfect conductors embedded in $SiO_2$ with following dimensions: 20 nm thick perfect conductors, 200 nm wide, and spaced 100 nm apart, as shown in Fig. 4(c). The source is positioned 0.3 μm above the particle. To demonstrate that resonance effects do not influence the simulations, we used a particle radius of R = 2.35 μm. The simulation results are presented in Fig. 4.

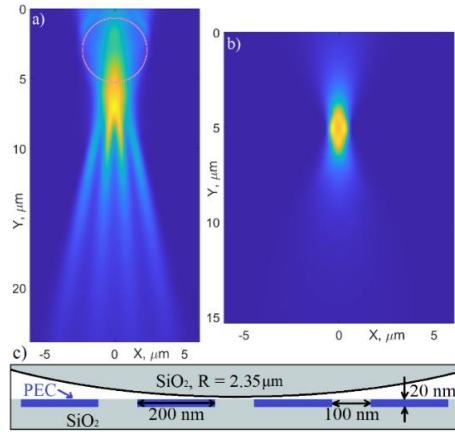

**Fig. 4.** Comparison of the image field in the reflection mode for the DVD disc for free space and microsphere. (a, b) Time-averaged image fields in reflection mode with and without the microparticle, respectively. Comparing distributions (a) and (b), we can conclude that the gaps are indistinguishable without the presence of the microparticle. For better contrast, we used $(I^{image})^2$. The source was 0.3 μm above the particle. (c) General calculation scheme: four perfect conductors, 200 nm wide, spaced 100 nm apart.

Next, we consider the image simulation for 50 nm pores in a gold-coated anodic aluminum oxide (AAO) membrane. Modeling the sample across its entire thickness is a cumbersome task, we limited the Al$_2$O$_3$ layer to 600 nm, with a PEC layer of 30 nm. The sample is divided into 50 nm pores with 50 nm spacing between them. The microparticle size is similar to that in the previous calculations, with a radius of R=3 μm. The radiation sources are placed 300 nm above the lower boundary of the AAO layer. The simulation results are presented in Fig. 5.

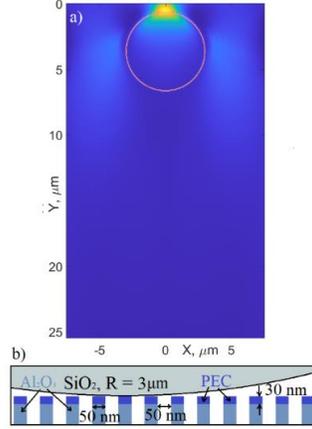

**Fig. 5.** Image field in transmission mode for the AAO membrane. (a) Square of the averaged image field, ($I^{image}$)$^2$; the red circle denotes the boundary of the particle. (b) Sample geometry: 50 nm pores with 50 nm spacing between them in a 600 nm thick Al$_2$O$_3$ substrate with a 30 nm PEC layer. The source is located 0.3 μm below the Al$_2$O$_3$ substrate.

As shown in Fig. 5(a), there are no pronounced maxima in the image field, leading us to conclude that, based on our model, 50-nm pores are indistinguishable. This discrepancy with experimental data may arise because our 2D model does not fully capture the complexities of a 3D experiment; additionally, resonance phenomena might be missed due to the limited accuracy of the FDTD method.

### 2.2 Laser source at 405 nm

To obtain a more reliable assessment, individual objects should be considered rather than periodic ones. We conducted similar calculations but focused on several individual objects. In studies [15, 16], individual objects were observed at distances on the order of 50 nm. Imaging was performed using a scanning laser confocal microscope operating at a wavelength of 405 nm. To maintain a fixed wavelength, continuous emission time signals were used instead of pulses. The simulation time was 1 ps for each source position.

According to that scheme, we simulated super-resolution for two perfectly conductive 120 nm wide dimers at a distance of 60 nm, as described in Ref. [15], and for three 136 nm wide particles at a distance of 25 nm, as described in Ref. [16]. We used reflection mode similar to the one used for the DVD disc mentioned above. The same material, SiO$_2$, was used for the microparticles. The obtained results are presented in Fig. 6. Both cases demonstrate super-resolution for a fixed wavelength of $\lambda = 405$ nm as observed in the experiment. The three particles (PEC objects) in Fig. 6(b) are barely distinguishable indicating the maximum resolution for this system. We also carried out the calculation in the absence of microparticles. In both cases, the structures were not distinguishable.

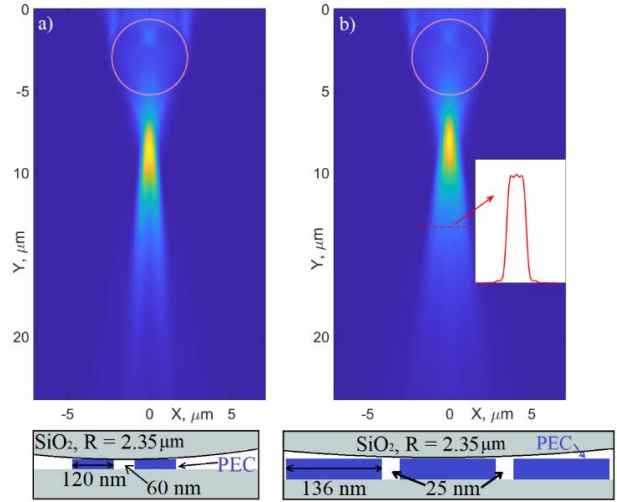

**Fig. 6.** Square of averaged image field in reflection mode with $\lambda = 405$ nm. (a) 120 nm wide dimers at a distance of 60 nm. (b) Three 136 nm wide particles at a distance of 25 nm. The inset panel shows the slice field at the dotted red line. The calculation geometry is shown in the lower panels. The particle height in both cases is 30 nm. The three particles (PEC objects) in (b) are barely distinguishable, indicating the maximum resolution for this system.

The optical resolution criterion $\lambda/2$ accepted in the literature refers to two point sources. The applicability of this criterion to real objects is widely discussed [8, 15], particularly regarding the distance should be compared between the edges of structures or their centers. If we consider a sample that is not separated by gaps, we can observe pronounced maxima, which are almost completely similar to the case with the gaps. For instance, the case of Fig. 6(b), in the absence of gaps, time averaging leads to these three maxima merging into one, while in the presence of gaps, these maxima remain distinguishable. The situation is different in the case of Fig. 6(a); even in the absence of a gap, two pronounced maxima can be observed, which do not disappear upon the averaging. Thus, although the two particles are distinguishable, in our calculations, they are indistinguishable from the case of a single particle without a gap.

## 3. Discussion

The presented calculations indicate that the source field and its spectrum are crucial parameters for the microparticle-assisted super-resolution and explain most of the phenomena observed in Ref. [1, 15, 16]. Our results provide convincing evidence of the distinguishability of structures with dimensions similar to Blu-ray DVD discs using a halogen lamp source. We demonstrate super-resolution on the order of 50 nm at the wavelength of $\lambda = 405$ nm, as in Ref. [15, 16]. Our calculations show that microparticle-assisted microscopy allows for the distinction of objects comparable in size to the wavelength of light, even when the distance between them is much smaller. However, the optical resolution may depend on the type of structure being considered. The spectrum of the used source contains wavelengths comparable to the size of the structures, allowing us to conclude that the sphere increases contrast, not resolution. To verify this statement, we performed calculations similar to that as in Fig. 6(a) with the same distance between particles but with 60 nm wide dimers. In this case, the particles are indistinguishable.

The presented investigation into dielectric microparticles using FDTD simulations confirms their ability to achieve super-resolution imaging well beyond the diffraction limit. This work not only demonstrates the practical viability of this technique for achieving resolutions down to 50 nm but also underscores the critical role of specific parameters such as source coherence and particle-substrate interactions. These insights pave the way for further exploration into more complex imaging scenarios and the development of next-generation optical imaging technologies with unprecedented resolution capabilities. Nonetheless, our model is still not complete. The 2D simulation does not allow for the simulation of more complex shapes, such as triangles or stars. In addition, we used ideal conductors to model the structures, that significantly reduced the overall modeling time. However, the results we presented can serve as a reliable foundation for further 3D modeling. Moreover, when using the FDTD method, we might have missed resonance solutions.

## 4. Materials and methods

### 4.1 Image field generation algorithm

Various algorithms for constructing an image field for harmonic fields are presented in Ref. [12, 13]. For our purposes, it is most appropriate to consider constructing the image field using the diffraction integral. By analogy with the three-dimensional case, for a two-dimensional TE field in the form $H_z(r,t) = H^s(r)e^{-i\omega t}$, the image field can be written as:

$$H^{im}(r_0) = \frac{i}{4}\int_\Gamma \left[ G(\mathbf{n},\nabla)H^{s*} - H^{s*}(\mathbf{n},\nabla)G \right] dl, \tag{2}$$

where $\Gamma$ is an arbitrary curve homotopic to an infinite line, $G = H^{(1)}(k|r-r_0|)$ is the Hankel function of the first kind, and * denotes complex conjugation, $k=2\pi/\lambda$, $\nabla = \partial/\partial \mathbf{r}$. The complex conjugation of the source field reverses the wave front, transforming the field from diverging to converging to the source. The $H^{im}(r_0)$ field does not depend on the type of the curve $\Gamma$, since the divergence of the integrand field is zero in the region where there are no sources.

Let us note an important property of the field $H^{im}(r_0)$. Let the source field propagates in the direction $y > 0$ and choose $\Gamma$ as the line $y = 0$. Then, the image field can be calculated only for the lower half-plane $y < 0$, since:

$$\frac{i}{4}\int_{y=0} \left[ G(\mathbf{n},\nabla)H^{s*} - H^{s*}(\mathbf{n},\nabla)G \right] dl = \begin{cases} H^{im}(r_0), y<0 \\ 0, y>0 \end{cases}. \tag{3}$$

The zero value in the half-plane $y > 0$ arises due to the Sommerfeld condition at infinity for the complex conjugate source field. Similarly, if the source field propagates in the direction $y < 0$, the $H^{im}(r_0)$ field can be calculated in the half-plane $y > 0$. Thus, field (2) is unidirectional and propagates strictly in the direction to the source.

Calculating field (2) in the harmonic case does not present any fundamental difficulties. How can this formula be applied to a pulsed field? Two approaches can be proposed. The first approach consists of decomposing the pulse into Fourier components and applying formula (2) to each harmonic. After using the inverse Fourier transform to obtain the image field components, the non-harmonic image field of the pulse can be reconstructed. This approach is quite cumbersome. Due to the linearity of Maxwell's equations and transformation (2), the image field can be calculated simultaneously for all harmonics. For this purpose, the source field for the image should be taken as the time-reversed source field. These sources should be arranged in such a way that Eq. (2) is satisfied for each harmonic.

In the FDTD algorithm, the source fields are specified in terms of field values at the nodes of the computational mesh. The image field (2) is the sum of two types of sources:

$$A: H^{(1)}(k|r-r_0|), \tag{4}$$

$$B: \frac{\partial}{\partial y} H^{(1)}(k|r-r_0|). \tag{5}$$

The source of type *A* is implemented by specifying the value of the derivative of the source field at a particular node of the computational grid. To implement source type *B*, it is necessary to set field values on neighboring nodes with alternating signs. Thus, to comply with formula (2), the source field must be specified along two adjacent lines.

The algorithm described above can be specified as follows. Let $H_z(x, y = y_o, t)$ and $H_z(x, y = y_o + dy, t)$ are source fields at the lines $y=y_o$, $y=y_o+dy$, accordinaly. Then the image field in these lines should be represented as:

$$H_z^{im}(x, y = y_o, t+dt) = H_z^{im}(x, y = y_o, t) + \\ + (H_z(x, y = y_o + dy, -t) - H_z(x, y = y_o, -t))/dy + H_z(x, y = y_o, -t)/dy, \quad (6)$$

$$H_z^{im}(x, y = y_o + dy, t+dt) = H_z^{im}(x, y = y_o + dy, t) - H_z(x, y = y_o, -t)/dy. \quad (7)$$

This approach retains all the properties of field (2). It is important to note that the sources are a so-called "soft source" where the new field value is added to the previous one and to the source field. This is necessary to eliminate the reflection of source fields from each other at mesh nodes. However, if invariance and unidirectionality are not required, specifying the sources can be significantly simplified. For instance, if $\Gamma$ is an infinite line, formula (1) can be rewritten as:

$$H^{im\prime} = \frac{-2i}{4} \int_\Gamma H^{s*} \frac{\partial G}{\partial y} dl. \quad (8)$$

The fields $H^{im\prime}$ and $H^{im}$ differ if the source field contains singularities. For example, in the case of a point delta source at the $r = r^{source}$, the following relation holds:

$$H^{im}(r) = H^{im\prime}(r) - i/4 G^*(r, r^{source}). \quad (9)$$

However, if the source is located at a significant distance from the plane $\Gamma$ ($\gg \lambda$), then the $G^*(r, r^{source})$ term can be neglected. In addition, if the image field of interest is also at a significant distance from the $\Gamma$ plane, we can write:

$$\frac{\partial G}{\partial y} \sim G \frac{y}{r}. \quad (10)$$

Thus, the field $H^{im}$ can be approximately calculated using the formula:

$$H^{im} \sim \int_\Gamma H^{s*} G \frac{y - y_o}{|r - r_o|} dl. \quad (11)$$

In this case, to specify the source for the image field, you can write:

$$H_z^{im}(x, y = y_o, t) = H_z(x, y = y_o, -t). \quad (12)$$

In this case, we used a "hard source" to suppress radiation from each point to the sides, with $y=y_o$ in formula (11). In our calculations, the source field for constructing the image was the field reflected or transmitted through the sample. In all cases, the distance to the time-reversal plane $\Gamma$ was several wavelengths or more (from 5 μm to 6 μm); for this reason, we used the simplified formula (12).

### 4.2 Image field generation for partial coherence source light

To simulate the image field, one must construct an appropriate source field. For the results involving the halogen lamp, we used spectral data from Ref. [2], as shown in Fig. 7, since the original work only indicated the peak wavelength. To simulate a broadband source field, we used time pulses with the corresponding spectrum shown in Fig. 7.

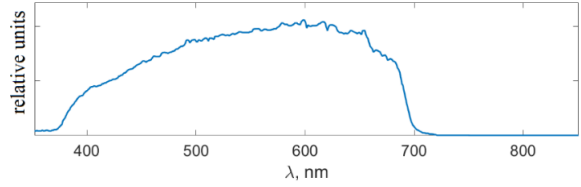

**Fig. 7.** The source spectrum used in our calculations for the halogen lamp [2].

The image field from each frequency in the spectrum should be summed by intensity. Since the source pulse contains time harmonics distributed according to the source spectrum, to find the total intensity, it is sufficient to integrate the field intensity over time, i.e.:

$$I^{image}(x, y, x_s) = \int_0^T |\mathbf{E}^{image}(x, y, x_s, t)|^2 dt = \frac{1}{2\pi}\int |\mathbf{E}^{image}(x, y, x_s, \omega)|^2 d\omega, \qquad (13)$$

here $T$ is the total simulation time, $x_s$ is the source coordinate. Due to the finite coherence radius, the image field from each section of the wavefront should also be summed based on intensity. The total image field observed in the experiment can be expressed as:

$$I^{image}(x, y) = \int I^{image}(x, y, x_s) dx_s \qquad (14)$$

Thus, the general scheme of our calculation is as follows. We define the source as a plane wave with a fixed width and spectrum according to Fig. 7. Next, the pulse transmitted or reflected through the microparticle and the sample is backpropagated through time into empty space according to formula (12). We then calculate the intensity using formula (13). Afterward, we shift the source by 100 nm and repeat the calculation. This process is continued as long as the fields contribute significantly. Finally, we sum the time-averaged intensities for each source position to obtain the final field, as given by formula (14).

To estimate the coherence radius of the field in the plane of the object, it is necessary to know the geometric parameters and the propagation characteristics of the field from the source. We chose a coherence radius of 600 nm based on its correspondence with the observed pattern in the experiment with 360 nm slits shown in Fig. 3. A larger radius leads to significant coherent effects in the absence of a microparticle, manifesting as pronounced maxima at the edges of the structure. A smaller radius results in a contradiction, as the gaps become distinguishable without the microparticle. In other calculations, we did not vary this parameter, but note that the results remain unchanged when it is reduced.

**Funding.** This work was supported by a grant from the Foundation for the Development of Theoretical Physics and Mathematics «BASIS». ZW thanks supports from Leverhulme Trust (RF-2022-659), Royal society (IEC\R2\202178) and Bangor University (BUIIA-S46910).
**Disclosures.** The authors declare no conflicts of interest.
**Data availability.** Data is available upon reasonable request.